\documentclass{article}
\usepackage{arxiv}
\usepackage[utf8]{inputenc} 
\usepackage[T1]{fontenc}    
\usepackage{hyperref}       
\usepackage{url}            
\usepackage{booktabs}       
\usepackage{amsfonts}       
\usepackage{nicefrac}       
\usepackage{microtype}      
\usepackage{lipsum}		
\usepackage{graphicx}
\usepackage{natbib}
\usepackage{doi}
\usepackage{url,hyperref,lineno,microtype,subcaption}
\usepackage[onehalfspacing]{setspace}
\usepackage{todonotes}
\usepackage{graphicx}
\usepackage{cleveref}
\usepackage{comment}
\usepackage{multirow}
\usepackage{diagbox}
\usepackage{array}
\usepackage{xcolor,colortbl}
\newcommand{\co}{CO\textsubscript{2}}
\newcommand{\eg}{e.\,g., }
\newcommand{\ie}{i.\,e., }
\newcommand{\wrt}{w.\,r.\,t.\ }
\newcommand{\cf}{{cf.\ }}
\newcommand{\HY}{\hyphenpenalty=25\exhyphenpenalty=25}
\newcolumntype{Z}{>{\HY\arraybackslash\hspace{0pt}}p{0.11\linewidth}}
\newcolumntype{M}{>{\HY\RaggedRight\arraybackslash\hspace{0pt}}p{0.11\linewidth}}
\newcolumntype{L}{>{\HY\RaggedRight\arraybackslash\hspace{0pt}}l}
\definecolor{greenminus}{RGB}{229,255,204}
\definecolor{greenequal}{RGB}{178,255,102}
\definecolor{greenplus}{RGB}{193,225,193}
\definecolor{orangeplusminus}{RGB}{254,227,128}
\definecolor{redminus}{RGB}{255,200,200}

\title{Climate Change \& Computer Audition:\\ A Call to Action and Overview on\\ Audio Intelligence to Help Save the Planet} 

\author{\textbf{Bj\"orn W.\ Schuller$^{1,2,3,*}$, Alican Akman$^1$, Yi Chang$^1$, Harry Coppock$^1$, Alexander Gebhard$^2$,} \\ \textbf{Alexander Kathan$^2$, Esther Rituerto-González$^{2,4}$, Andreas Triantafyllopoulos$^2$, Florian B. Pokorny$^{2,5}$} \\
              \AND
                \\$^1$GLAM -- Group on Language, Audio, \& Music, Imperial College London, UK \\
                $^2$EIHW -- Chair of Embedded Intelligence for Health Care and Wellbeing, University of Augsburg, Germany \\
            	$^3$audEERING GmbH, Gilching, Germany \\
            	$^4$GPM -- Group of Multimedia Processing, University Carlos III of Madrid, Spain \\
            	$^5$Division of Phoniatrics \& Division of Physiology, Medical University of Graz, Austria
              \And
               \texttt{schuller@ieee.org}
               }
\renewcommand{\shorttitle}{\textit{Climate Change \& Computer Audition}}
\hypersetup{
pdftitle={Climate Change \& Computer Audition},
pdfauthor={Schuller {et~al.}},
pdfkeywords={Computer audition, audio intelligence, climate change, earth, water, air, fire, environment, call to action},
}

\begin{document}
\maketitle

\begin{abstract}
Among the seventeen Sustainable Development Goals (SDGs) proposed within the 2030 Agenda and adopted by all the United Nations member states, the 13$^{th}$ SDG is a call for action to combat climate change for a better world. 
In this work, we provide an overview of areas in which audio intelligence -- a powerful but in this context so far hardly considered technology -- can contribute to overcome climate-related challenges. 
We categorise potential computer audition applications according to the five elements of earth, water, air, fire, and aether, proposed by the ancient Greeks in their five element theory; this categorisation serves as a framework to discuss computer audition in relation to different ecological aspects.
Earth and water are concerned with the early detection of environmental changes and, thus, with the protection of humans and animals, as well as the monitoring of land and aquatic organisms. 
Aerial audio is used to monitor and obtain information about bird and insect populations. 
Furthermore, acoustic measures can deliver relevant information for the monitoring and forecasting of weather and other meteorological phenomena. 
The fourth considered element is fire. 
Due to the burning of fossil fuels, the resulting increase in \co\ emissions and the associated rise in temperature, fire is used as a symbol for man-made climate change and in this context includes the monitoring of noise pollution, machines, as well as the early detection of wildfires. 
In all these areas, computer audition can help counteract climate change. 
Aether then corresponds to the technology itself that makes this possible. 
This work explores these areas and discusses potential applications, while positioning computer audition in relation to methodological alternatives.
\end{abstract}


\section{Introduction}
\label{sec:introduction}
Our climate is rapidly changing.
According to the 2021 Assessment Report of the International Panel for Climate Change (IPCC): ``Human influence has warmed the climate at a rate that is unprecedented in at least the last 2000 years"~\citep{ipcc2021}.
\co\ emissions caused by human activities are a major driver of those changes, and have caused a dramatic rise in temperatures from the mid 19\textsuperscript{th} century to the present day~\citep{ipcc2021}.
This rise in temperature has impacted the environment in various ways: increased precipitation, rise of sea levels, loss of glacier mass, desertification, heatwaves, and an increased frequency of extreme weather events~\citep{ipcc2021}, with those changes in turn leading to a massive loss of natural habitats and biodiversity~\citep{hoegh2010impact, ceballos2017biological}.
These extraordinary circumstances are also detrimental to human well-being and health~\citep{watts2015health}, compromise food security~\citep{wheeler2013climate}, and lead to increased armed conflicts~\citep{raleigh2007climate}.

Accordingly, climate change has been described by many as `the greatest challenge of our time', calling for an equally outstanding response by the international community.
Efforts at governmental and institutional level to curb \co\ emissions and limit environmental pollution have largely dominated public conversations, and rightfully so, as a coordinated push towards transforming our societies is paramount to mitigating the more catastrophic effects of climate change.
To that end, major focus has been placed on regulating polluting activities. In many respects, this can be seen as a move to limit the use of certain polluting technologies across several facets of human activity.
Future technological breakthrough, however, is expected to play a major role in curbing, and even reversing, emissions, for example through the development of renewable energy sources~\citep{panwar2011role} and carbon sequestration technology~\citep{figueroa2008advances}.

Artificial intelligence (AI) is one of those technological paradigms with a great potential for assisting the fight against ecological catastrophe~\citep{cowls2021ai}.
Its major promise comes from the capacity to analyse vast amounts of diverse, multi-dimensional data and the ability to enable distributed, localised interventions in ecosystems in need.
AI can be used in two major ways with respect to climate change: to \emph{understand} it and to \emph{combat} it.
\emph{Understanding} corresponds to the detection and categorisation of patterns; this could be (global or local) weather patterns, the tracking of specific environmental indicators, or the monitoring of animal populations.
\emph{Combating}, in turn, aspires to utilise AI-powered technologies to counteract specific changes.
This differentiation is consistent with the common categorisation of AI systems into those that \emph{passively observe} and those that \emph{actively interact} with their environment.

Computer audition (CA) is the sub-field of AI that encompasses all facets of the auditory information stream; it can thus function both as a passive perception system and an active, audio-generating agent.
Even though CA plays an important role within the AI community, its potential for helping combat climate change remains largely untapped.
The aim of this article now is to highlight application areas where CA can be utilised in this context.

\begin{figure}
  \includegraphics[width=\linewidth]{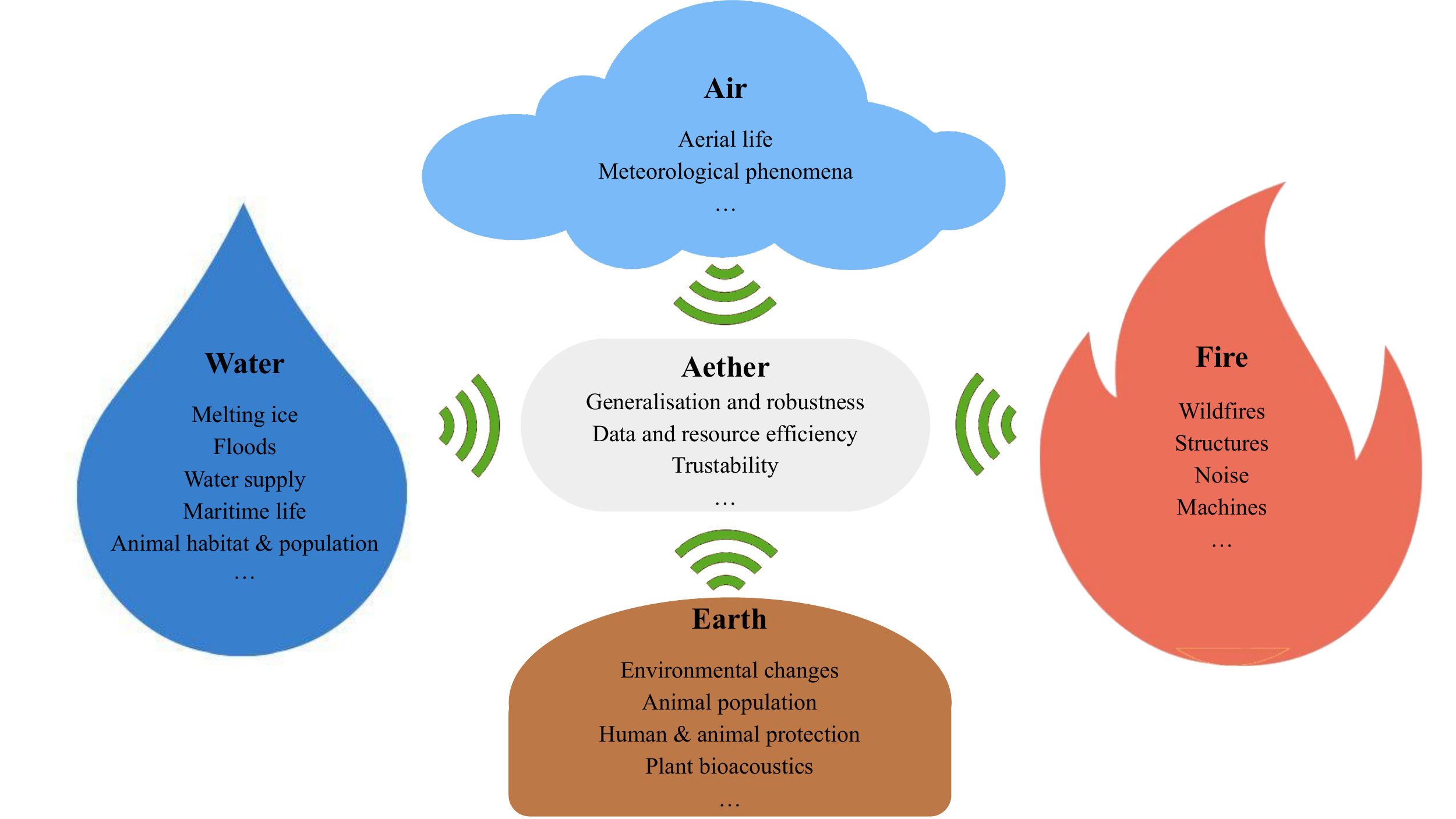}
  \caption{Overview of the different areas where computer audition can deliver value to save the planet.}
  \label{fig:elements}
\end{figure}

The catastrophic effects of climate change are spread over different parts of our planet's natural ecosystem, the diverse characteristics of which warrant for specialised considerations for any technology that purports to be utilised in the fight to protect them.
In our attempt to categorise the different application areas for which CA can be useful, we adopt the four element theory proposed in Ancient Greek and Persian mythology~\citep{partington1989short}: \emph{earth}, \emph{water}, \emph{air}, and \emph{fire}, a quick overview of which is provided in \cref{fig:elements}.
\emph{Earth} and \emph{water} concern themselves with terrestrial and aquatic ecosystems.
\emph{Air} stands for applications targeting gas monitoring, bird and insect population monitoring -- in a nutshell, everything that exists in our planet's atmosphere.
\emph{Fire} is used in both a metaphorical and a literal sense: as the biggest threat to our environment is the rise of temperature caused by \co\ emissions, starting with the invention of the internal combustion engine, we use \emph{fire} as a symbol for all those human activities that threaten our environment; but (wild)fire in itself is also a big threat to earthly ecosystems, so we also accordingly analyse how CA can assist in its detection and mitigation.

This leaves \emph{aether}, a fifth element proposed by Aristotle which constitutes the `essence of things', as a symbol for AI, the enabling technology making a variety of innovative approaches possible.
The word itself calls attention to the popular depiction of AI-powered systems as magical solutions for several types of problems -- a fallacy we attempt to avoid by sketching out extrapolations of existing, well-founded approaches, that may lead to the realisation of the promise of CA to assist in climate change mitigation, as well as highlighting associated challenges.

\begin{figure}
  \includegraphics[width=\linewidth,trim={1cm 1cm 1cm 1cm}]{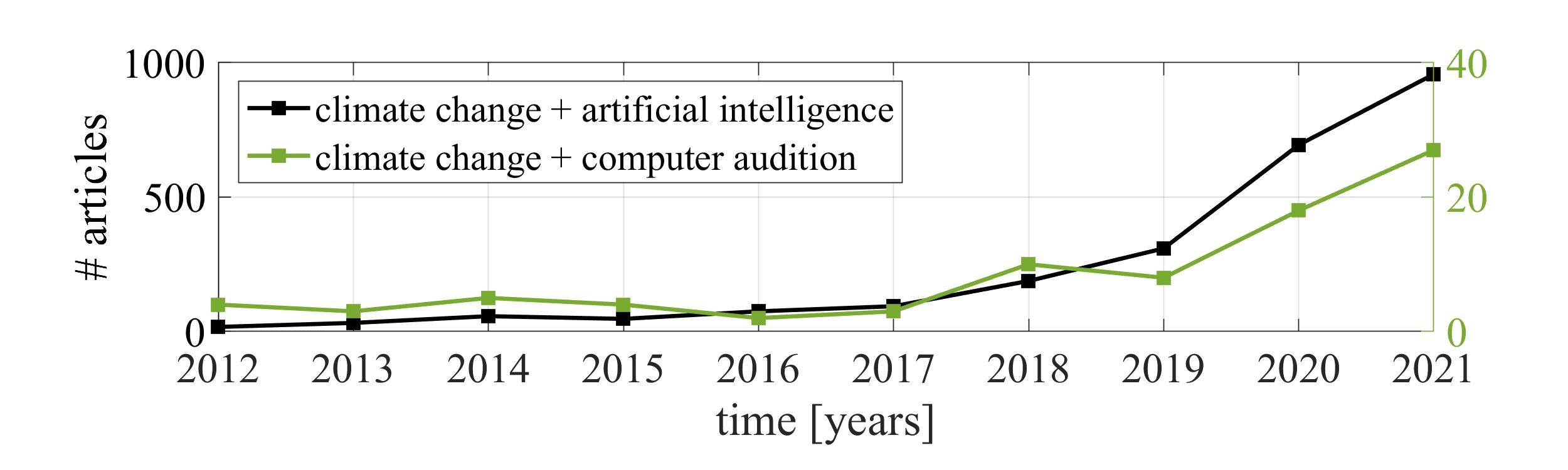}
  \caption{Number of (\#) articles indexed in the Web of Science Core Collection over the last 10 years (as of 21 December 2021) related to climate change and artificial intelligence (black; left y-axis) vs climate change and computer audition (green; right y-axis).}
  \label{fig:literature}
\end{figure}

To date (as of 21 December 2021), there are 321\,776 articles indexed in the Web Of Science Core Collection related to climate change (search term: ``climate change''), with its number having triplicated over the last 10 years (13\,678 in 2012 vs 39\,595 in 2021). 2\,571 of these articles deal with AI (search term: ``climate change'' AND (``artificial intelligence'' OR ``machine learning'' OR ``deep learning'')) and were to a large part (76\,\%) published in the last three years. With a total amount of just 89 Web of Science Core Collection indexed articles (search term: ``climate change'' AND (``artificial intelligence'' OR ``machine learning'' OR ``deep learning'') AND (audio OR acoustic* OR sound OR noise)), CA has played a minor role in research on climate change so far. However, same as for AI in general, there has also been a significant increase of CA-related articles in context of the climate change in recent years. 45 of the 89 articles were published in 2020 and 2021 (see \cref{fig:literature}). 

These numbers indicate that CA represents a relatively underexplored, but promising field for fighting climate change.

In the following, we give a comprehensive, but non-exhaustive overview of important recent work using CA in order to assist our planet. For this overview, identified relevant studies were categorised according to application areas. In this regard, each of the following  sections is dedicated to one application area represented by the previously introduced elements earth, water, air, and fire, followed by more general AI-related aspects being addressed in a section on aether. As the opposite of recording and automatically analysing audio, we then also outline the idea of intelligent audio generation to help the ecosystem, in a separate section. Finally, we discuss the advantages and limitations of CA compared to alternative methodology at hand to help the environment; we address relevant ethical issues, and close by providing an outlook of what has been reviewed in this paper.

\section{Earth}
\label{sec:earth}
This section deals with the first element, \emph{earth}, and discusses the potential CA has to combat climate change in this setting. In doing so, we focus on the recognition of changes
on earth at an early stage~\citep{martin2008impact}, which is of great importance  in order to avert potential damage, and therefore to be able to react in good time and to protect living beings accordingly.

\subsection{Environmental Changes}
The detection of environmental changes is a crucial component of future challenges~\citep{martin2008impact}. As these environmental changes are becoming increasingly rapid and have an impact on many areas of life, it is important that these changes are identified at an early stage so that appropriate countermeasures can be taken. Environmental changes can be natural changes, such as natural disasters in the form of earthquakes or volcanic eruptions~\citep{banholzer2014impact,hallegatte2016natural}, but also man-induced changes~\citep{hansen2012paleoclimate}, like deforestation. Both types of changes must be recognised at an early stage so that appropriate mitigation strategies can be planned. 

While previous methods for environmental change detection primarily use visual data~\citep{kim2019deep,liu2016application}, CA offers the possibility to use a new, so far largely untapped, information channel. With this new possibility of hearing environmental changes, several sounds can be perceived. One category are man-induced changes~\citep{hansen2012paleoclimate}, such as deforestation of the rainforest. In this case, \eg the detection of chainsaws or large machines can serve as an indicator. Another category is the detection of natural events~\citep{banholzer2014impact,hallegatte2016natural}: environmental changes, such as earthquakes, cause perceptible sounds to various natural objects, which can also be heard with special devices~\citep{muller2015acoustics}.

\subsection{Animal Population}
CA was already used to classify several animals in the past~\citep{gibb2019emerging,stowell2019birdchallenge,schuller2021interspeech}. The availability of audio recordings over a wide area could not only identify which animals are present in the region, but also roughly how many of them are present and where they are moving. Animal monitoring with CA can therefore be a great opportunity~\citep{gibb2019emerging}. Big changes in the movement of animals could, \eg indicate possible dangers such as natural disasters~\citep{garstang2017understanding}. All of this can be detected without having seen a single image of the actual danger, but only with the audible movement profile of the animals. 

By detecting which and how many animals are at a given location and where they are moving to, animal movement profiles can also be used to analyse their population dynamics at different locations. If the population of certain animals at a location decreases sharply, this can be used as an early warning system to take countermeasures in time. This not only provides insight into the evolution of species in different locations, but also the opportunity to save endangered species from extinction through early interventions.

In addition, animals produce diverse sounds in nature, such as when defending territories, fighting with other animals, or during mating season~\citep{blumstein2011acoustic}. With the help of these sounds, biologists can gain a variety of valuable insights into the animals' way of life and use this information for further analyses. Moreover, the automatic characterisation of biological traits in animals, such as by sex and age groups, can help monitor the health of packs and herds in the wild~\citep{blumstein2011acoustic}.

Another major problem of our time is the change in biodiversity and species extinction. This is again an area where CA can add value~\citep{gibb2019emerging}. It is a great advantage to be able to cover extensive areas with microphones to gain insights over animal population. Audio recordings are particularly useful in places where it is difficult to make full-coverage camera recordings, such as in forests. Furthermore, the presence and composition of different acoustic sounds provides information not only about the composition of vocalising species and the orthopteran species richness, but also about the structure of the landscape and the intensity of land use~\citep{muller2022land}.

\subsection{Animal Protection}
In addition to the previous section, not only the movement profile of animals can be used as an indicator for dangerous changes, but this information can also be combined with directly perceptible sounds, such as machinery sounds which indicate the presence of humans. Therefore, such a fully comprehensive audio understanding allows early detection of environmental changes and thus danger for the animals that might otherwise have gone undetected. 

Nature is not the only source of danger to animals; they can also pose a danger to one another or danger is posed by human beings. For example, poachers pose a threat to wild animals and predators to farm animals, such as livestock or poultry. CA can be of benefit here as well, such as by recognition of gunshots \citep{Turian22-H2H}. 

In agriculture, attack by wild animals is often a major hazard. Several solutions have therefore already been developed in the field of computer vision to detect relevant situations~\citep{andavarapu2017wild}. Nevertheless, visual solutions have the problem that cameras usually cannot fully cover large pastures. In contrast, CA offers the advantage that audio recordings of large pastures can be fully recorded with only a few microphones and are therefore much more suitable for practical use cases. These audio recordings can then be used to determine which animals are in the pasture. In case of an attack on the herd by wild animals, this knowledge can be used to inform the farmer accordingly in time. An attack by humans on animals, e.\,g., by poachers, can also be heard and detected at an early stage, thus, endangered animals can be protected from other animals as well as from humans~\citep{kamminga2018poaching}.

In this respect, the recognition of emotion and social (and more general cognitive, phsyical, and health) state of animals by audio \citep{Hantke18-WIM} can also help monitor their health and wellbeing or recognise impeding threats. 

\subsection{Plant Bioacoustics}
On a different note, the field of plant bioacoustics is a rising exploration field which focuses on the sound waves created by plants and insects. Bioacoustic tools have been applied to measure mechanical properties of plant structures, optimise mechanical harvesting, and detect the distribution of root systems, as well as to monitor plant health, photosynthesis, and ecology  \citep{plants_stress_2019}. Recent experimental studies open the possibility of assessing the stress of plants by using machine learning algorithms on acoustic signals, \eg analysis of ultrasound emitted by plants to determine their health~\citep{bioacoustic_tools_2018}.

In addition, acoustic and vibration sensors are used by entomologists to detect hidden infestations of invasive insect species, and to monitor insect movement, feeding, and mating activities on host plants \citep{behavioural_ecology_2005}. Novel studies \citep{bioacoustic_tools_2018} consider the use of bioacoustic tools to analyse plant health and structural characteristics, and discuss how combinations of spectral-, temporal-, and spatial-distribution features of signals detected in plants can be interpreted in ways that properly enable reliable assessment of hidden pest infestations, including invasive insect species of importance for plant biosecurity.

The acoustic performance of some organisms show the magnification of the effects of climate change. Drought, for example, produces stress on trees and leads to an increased vulnerability to insect attacks. Insects are drawn to stressed trees using chemical signals, but also are attracted by the sounds emitted by tree cells  \citep{insect_trees_2016}. These sounds, which are produced by forest trees when being under drought stress, are known as \emph{cavitation}, which is the result of cells collapsing by gradual dehydration. The majority of these sounds emitted are within a frequency range of 20\,kHz to 200\,kHz \citep{plants_sounds_2012} and carry information for insects, that can perceive such signals. 

Defoliating insects have a large impact on ecosystems and are influenced by climate change as well~\citep{pureswaran2018forest}. Therefore, changes of their behaviour can be used as an indicator of \eg an increasing amount of \co\ in the atmosphere. Early detection of these changes is of great importance. Thus, this acoustic feedback from insects can have a positive effect, as it can be detected by CA and thus indicate defoliation, forest decline, and \co\ increase. Insects will be addressed in more detail in the section on \emph{air}.

\section{Water}
\label{sec:water}
\emph{Water} is often referred to as the `element of life'. It serves as nutritional source \citep{popkin2010water,ritz2005importance,self1992nitrates}, as a habitat for various animals \citep{alava2012habitat,samuel2005underwater,freeman2013coral} or can simply be used to maintain hygiene and therefore avoid several diseases \citep{mara2003water,ashbolt2004microbial,pengpid2012hygiene,curtis2007dirt,matta2017health}. That is, every living being on our planet needs water to survive. 
Therefore, we have to ensure the quality and quantity of this vital source of life. 

Water can be encountered in three physical states: solid, liquid, and gas. Thus, we have by default various ways to perceive this element. For instance, considering its solid state, ice, CA could hear the cracking ice long before we could observe it melting visually, e.\,g., when chunks of ice break down from icebergs or glaciers are disappearing; which might be indications of an increasing temperature in the environment. CA also has the potential to detect possible hazards like floods or tsunamis early on or to monitor the water supply of an area. Besides, coral reefs, which are often called the ``rainforests of the sea''~\citep{knowlton2001future}, and its biodiversity can be tracked. Regarding animal observation, the animal communication and movements can be investigated quite easily with CA. It is much more efficient to record related sounds with microphones under the water surface covering a larger area than, for example, diving in with cameras. In water, sound propagates with a higher velocity and over greater distances than in air. This is conversely to vision, which is dramatically hindered in water as compared to travelling through air.
In the following subsections, these various application areas will be described in more detail.

\subsection{Melting Ice}
In times of melting polar ice caps it would also be helpful to monitor the process of the melting ice. That is, we can hear for cracking sounds in icebergs, ice floes or glaciers ~\citep{urick1971noise,ashokan2016ice,lee2013underwater}. If there is an accumulation of cracking noises or a huge explosive crack it might be a clue for the melting of ice which is affected tremendously by increased temperatures and, thus, an indicator of climate change. On the one hand, the detection of melting ice can be important for study purposes, e.\,g., to investigate how long the melting process has lasted before a chunk of ice breaks down from a glacier \citep{deane2019underwater}. On the other hand, it can be utilised for the early prediction of avalanches or floods and, therefore, enable precautions or appropriate countermeasures. In the best case, catastrophes could even be prevented.

Moreover, there can be major landslides due to melting glaciers causing enormous landslides and tsunamis \citep{marchenko2012tsunami,higman20182015}. According to researchers, there is a ticking time bomb at the moment in Alaska within the \textit{Barry Arm} area, the Barry glacier, which has the potential of causing a mega-tsunami \citep{dai2020detection}. Moreover, the risk of major landslides due to melting glaciers and, thus, the risk of arctic mega-tsunamis is rising steadily \citep{schiermeier2017huge}.

\subsection{Floods}
But not only the sound of cracking ice might be helpful in predicting floods or flood waves, but also the sound of flowing water (e.\,g., in rivers), since this sound reflects the water flow velocity \citep{tonolla2009flume,lumsdon2018soundpeaking}. The sound is generally generated by particle collisions through streamed sediment movements as well as the flow of water over submerged obstructions \citep{lumsdon2018soundpeaking}.
These factors might be very helpful for roughly inferring the amount of flowing water and predicting overflowing rivers and lakes in order to appropriately prepare agricultural land and the surrounding civilisation for such scenarios. In this connection, hydrophones -- underwater microphones -- could be placed at locations, which have a high potential of being hit by floods such as estuaries at glaciers, water reservoirs, or moors. 

\subsection{Water Supply}
In addition to the early prediction of floods due to the sound of running water we can also predict the opposite. That is, if there is minimal audible sound it might be a clue for drying up rivers, lakes or natural fountains, and thus be a sign of water shortage in a certain region. Via early prediction of water scarcity in specific areas, artificial irrigation facilities could be constructed in advance or the people living in such areas could be relocated. 

However, water supply does not only apply to overall regions but also to individual beings.
To that end, we can exploit the fact that all living beings contain water in their bodies.
For instance, based on ultrasonic sounds of trees, their water supply could be assessed and conclusions about their water supply could be drawn \citep{ponomarenko2014ultrasonic}.

\subsection{Maritime Life}
Soundscapes have been used in terrestrial landscapes and restoration, and are recently being used to monitor coral reefs underwater \citep{lin2021exploring}. In places where dynamite fishing is common practice, the reefs end up being destroyed \citep{slade2015dynamite,wells2009dynamite}. 
Such endangering human activity can of course be automatically heard. 
In addition, due to rising sea temperatures and the acidification of the oceans, mass `bleachings' of coral reefs is now a common occurrence.
A healthy coral reef can be distinguished through the acoustic modality, as their bright, loud, and diverse soundscapes guide the recruitment of reef organisms, but those disappear when damage occurs \citep{coral_reef_2021}. 
Restoring the reefs can bring those sounds back to life, as well as the ecosystem \citep{coral_reef_2021}.  Besides, the acoustic enrichment consequence of generating and emitting sounds from such healthy soundscapes underwater can enhance fish community development on degraded coral-reef habitats \citep{coral_reef_2019}.

By sensing underwater animal sounds, e.\,g., produced for communication purposes or caused by movement patterns, conclusions can be drawn \wrt their population, behaviour, and habitat~\citep{clark1984sounds,cummings1987sounds,klinck2012seasonal,Schuller19-TI2}. Large bioacoustic achrives like the \textit{Orchive}~\citep{ness2014human,ness2013orchive} are very useful in this context~\citep{bergler2019orca}. 

The aforementioned scenarios can be significantly influenced by changing environmental conditions. Therefore, they are key factors in which to measure the repercussions of climate change, fight its consequences, and tackle the damage caused by humans. 

\section{Air}
\label{sec:air}
The mass of air that we denominate the atmosphere represents approximately only the 5\,\% of the total volume of our planet, but it is crucial for all forms of life in it. CA can make a big impact in the sustainment of both natural and human-made environments through air as it is -- from a human perspective -- the common medium for the propagation of soundwaves.

In the wild, the atmosphere supports a thriving ecosystem for birds, bugs, and bacteria. Acoustics can have a big impact in habitat monitoring by tracking birds migration and insects populations. Audio information can also be used for the efficient administration of crops by controlling and monitoring pests. Besides, acoustic measures can give us relevant information for weather and meteorological effects monitoring and forecasting, such as air pressure measures and wind characterisation \citep{mars_wind_2021}. Gathered data can be used from the predication of overcast and sunlight ratios for enhancing agriculture production sustainability, to the prediction of natural disasters such as tornadoes \citep{tornado_oklahoma_2018}. 
In this section, we will discuss more in depth, where CA in connection with air can make a contribution to save the planet.

\subsection{Aerial Life}
The atmosphere and the earth's ecosystems are parts of a coupled system. The disciplines of aerobiology and aeroecology explore how animals, plants, and other organisms live in, move through, and interact with the atmosphere. 

Birds are rapidly affected by changes in the environment. In the mining industry, caged canaries were carried down by miners into the mine tunnels. Whenever there was a leak of dangerous gases, such as carbon monoxide, the gases would kill the canaries, which served as a warning for the miners to exit the tunnels immediately. Birds are messengers that tell us about the health of the planet because they are widespread, they connect habitats, resources and biological processes. They also contribute to ecosystem services -- as natural enemies of pests, pollinators of fruit, and seed transporters \citep{birds_services_2008}. Birds also play a key role in cycling nutrients and helping to fertilise marine ecosystems \citep{bird_droppings_2020}. They are thus crucial for the sustainability of the environment.

Whether ecosystems are managed for agricultural production, wildlife or water, success can be measured by the health of birds. Bird sound classification aids to determine their presence, tracking their migrations, and measuring their population \citep{rajeev2019Birdbioacoustics,stowell2019birdchallenge}.
A decline in bird numbers informs about a damaged environment \citep{birds_decline_2004}, \eg due to habitat fragmentation and destruction, pollution, and pesticides introduced species, etc. Birds provide insect and rodent control, which results in tangible benefits to people. Insect outbreaks can annually have a huge negative economic impact in agricultural and forest products, and some birds can be effective to substantially reduce insect pest populations without the health, environmental, and economic risks of harmful pesticides \citep{state_of_birds_2009}.
Microphone arrays can give us bird data in a continuous 
manner, where other sensors such as cameras would struggle to monitor such large areas. 
For instance, for the tracking of nocturnally migrating birds \citep{nocturnal_birds_2010}, for which the use of night vision cameras -- requiring complex manufacture with high-voltage power supplies to operate -- would be expensive. The acoustic performance of bird communities reaches its maximum at dawn and dusk, when species are contemporarily singing and producing choruses \citep{bird_choruses_2015}. Measuring the length, energy,  and frequency components of choruses can, \eg give insight about the ambient temperature \citep{sparrows_2016}, as ambient temperature can cause changes in the physiology of organisms. 

\subsection{Insects and Pests}
In addition to birds, insects are the other large group of species that populate the air. They are under immense pressure from land use intensification and climate change effects, threatened with extinction or showing significant population declines \citep{birds_extinction_2012}. Insects are essential in food chains and cycles; they pollinate fruits, flowers, and vegetables, and are also very important as primary or secondary decomposers. Many insects are omnivorous, and eat a variety of foods including plants, fungi, dead animals, and decaying organic matter, helping breaking down and disposing wastes. Predatory or parasitic insects help keep pest populations, such as insects or weeds, at a tolerable level. They are also the sole food source for many amphibians, reptiles, birds, and mammals. 

Especially, bees contribute to complex, interconnected ecosystems that allow a diverse number of different species to co-exist \citep{bees_environment_2020}. Acoustical non-intrusive sensors are being introduced along with temperature and moisture sensing for bee hive colony activity health and status monitoring, for its subsequent analysis using machine learning \citep{beehives_2019}.

In recent decades, acoustic devices have provided nondestructive, remote, automated detection and monitoring of insect and pest infestations for pest managers, regulators, and researchers \citep{insects_review_2011}. Microphones are useful sensors for airborne signals, specially ultrasonic sensors, which are particularly effective for detecting wood-boring pests like termites at frequencies of more than 20\,kHz.

Insect pests can also pose a serious threat to agricultural and forest ecosystems, but the difficulty to control insect pests makes them challenging to prevent. Novel research on methods for acoustic data analysis based on active sound production by larvae (\ie \emph{stridulations}) can give insight into larval ecology produced by pests \citep{larvae_2019} and opens up a new road for pest control.
This acoustic monitoring of larvae, and the data analysis for automatically detecting audio sections with stridulations, can provide an estimate of their activity, enabling non-invasive species-specific pest monitoring.

As reliability and ease of use increase and costs decrease, acoustic devices have considerable future promise as cryptic insect detection and monitoring tools.

\subsection{Meteorological Phenomena}
The movement of winds have huge implications for storm systems and precipitation patterns. Specifically, westerlies winds transport dust from desert regions to faraway locations, making changes in the environment \citep{dust_winds_2021}.
Recording aeroacoustic noise generated by wind flowing past a microphone can provide wind speed and wind orientation. A recent study analysed the frequencies composing the wind-induced acoustic signal measured by microphones \citep{mars_wind_2021}. The acoustic spectra recorded under a wind flow can be decomposed in the low-frequency range, mainly reflecting the wind velocity, and the higher frequency range, regarded to depend on the wind direction relative to the microphone. Therefore, microphones as tools to monitor the wind have a huge potential to show climate disruptions and potentially help adaptively control energy-generating wind mill farms, while also hearing potential disruptions in their routine.

On a different note, numerous geophysical and anthropogenic events emit acoustic waves below the human hearing range of about 20\,Hz, \ie infrasound, including hurricanes and tornadoes. The rate of increase of severe storm environments becomes greater in the northern hemisphere due to temperatures rises \citep{tornado_climate_2021}. Tornado-producing storm systems emit infrasound up to 2 hours before tornado genesis. This can be detected from large distances (in excess of 150\,km) due to weak atmospheric attenuation at these frequencies. Thus, infrasound could be used for long-range, passive monitoring and detection of tornado genesis as well as characterisation of tornado properties \citep{tornado_oklahoma_2018}.

\section{Fire}
\label{sec:fire}
Perhaps no other element better symbolises our current predicament than \emph{fire}, as the burning of fossil fuels has been largely responsible for the rise in \co\ emissions and the resulting temperature increase.
Thus, fire in our work primarily stands for man-made climate change.

However, fire itself is also a consequence of man-made climate change. 
The world's flora and fauna are under threat from the increased frequency and strength of forest fires.
Tragically, those fires are now increasingly affecting residential areas, causing immense damage to communities.
Therefore, we will additionally concern ourselves with the early detection of fire itself, where CA can also prove a valuable asset. 

In this section, we will discuss the usefulness of CA in monitoring those aspects of human behaviour that are most detrimental to the environment.

\subsection{Wildfires}
Fires are primarily detected using either images or sensors for temperature, humidity, and smoke.
While those measurements provide the de-facto standard for the detection of forest fires, they cannot readily distinguish between different types of fires.
However, this differentiation is crucial in a world where wildfire seasons are getting longer and more intense, and fire brigades need to figure out how to best spread their limited resources.
In that respect, crown fires, which burn through the upper layers of trees, are more intense and have a higher velocity than surface or ground fires.
Distinguishing between these different types of fires is crucial for combating large wildfires, as it determines the type of response needed.
While other modalities, such as imagery sensors, can be utilised for this categorisation, none of them is adequate on its own as each comes with its shortcomings: Cameras can be limited by smoke, while temperature, humidity, and smoke are best tailored to detect the presence of fire, but might be insensitive to its type.
To that end, the acoustic properties of fires have been previously shown to vary across different types, thus, enabling their classification based on auditory perception~\citep{khamukhin2016spectral, khamukhin2017algorithm}.
Furthermore, microphone arrays can be used to determine the location of lightning thunder sources, since lightning strikes are one of the major causes of forest fires.
This lays out a new promising direction for the timely classification of different forest fires which could provide crucial information for early combating them.

A further complicating factor is that fires will naturally disrupt any monitoring system put in place to detect them; \eg as the fire itself, or the water used to extinguish it, destroys the sensors.
Thus, immediately after the fire, affected areas will be left without sensory coverage.
This constitutes a crucial challenge as the re-ignition of existing fires in already burnt-out areas is a major problem for firefighters.
A poignant example is the August 2021 fire of Varympompi (near Athens, Greece)\footnote{https://go.ifrc.org/reports/14615}, which was initially controlled by the fire brigade, only to be re-ignited a few hours later due to insufficient supervision, with the majority of its destruction coming with the second wave.
This illustrates that it is imperative to rapidly (re-)deploy sensing equipment in the affected areas.
Those areas, however, will be heavily affected by smoke (especially if the fire is still ongoing in nearby land), thus, making it harder for image or smoke sensors to detect potential sources of rekindling.
This necessitates the use of sensors whose effectiveness is not inhibited by the presence of fire and that is tailored to the classification of its type, rather than its detection only -- another potential advantage of auditory perception over other modalities.
However, domain adaptation is going to be a problem as the sound of fire will differ between forests.

\subsection{Structures}
Communities around the world struggle to reconvene their lives in the aftermath of a catastrophic fire, especially if the fire  affected residential areas.
One of the usual reaction to such fires is the promise to rebuild all destroyed or affected buildings.
Unfortunately, building is one of a major source of \co\ emissions, necessitating an environmentally-friendly rebuilding paradigm.
This includes the effort to salvage as much as possible from the remnants of an urban fire.

A major consideration after a building fire is its effect on structural integrity, which is the ability of a structure to withstand the required load without collapsing.
If this integrity is compromised to an irreparable extent, then the building needs to be demolished and reconstructed.
Determining the extent of the damage, however, is not an easy feat, especially as any investigation needs to be conducted by means of non-invasive techniques that do not further compromise a building's integrity.
This integrity is dependent on the change of strength of the materials a building is made from, which is in turn reflected by the way sound propagates; thus, CA presents a novel avenue of investigating changes in material strength which can provide useful information on the damage a building has sustained.
As a recent example of such work, \citet{schabowicz2019identification} study the condition of materials subjected to fire, and utilise acoustics for identifying the degree of degradation of fibre-cement boards.

\subsection{Noise}
\label{subsec:noise}
From primitive people's first use of fire to modern day industrial revolution, the shadow of fire can be felt with increasing frequency.
Its presence is not only felt through visual or temperature stimuli, but through auditory stimuli as well.
In particular, the latter are another byproduct of modern industry, with detrimental effects to human well-being and ecological equilibrium.
Industry, machinery, and cities are the main sources of acoustic contamination. 
This has a major environmental impact and significant detrimental effects on wildlife~\citep{halfwerk2015pollution, anthropogenic_noise_2019, CHAN20111}, generating stress and jeopardising wildlife reproduction~\citep{ALQUEZAR2019163}, potentially reducing biodiversity~\citep{sordello:hal-02948589}, affecting wildlife communication strategies~\citep{wildlifecom}, and even contributing to the outright extinction of some species~\citep{templeton2016traffic, animal_extinction_2018}.
With CA, abnormal noises can be monitored, identified, and dealt with in timely and decentralised fashion, as sensors can be spread across areas of interest.
These could include wildlife habitats near industrial zones, motorways, airports, or railways; thus, zones where the presence of nature and mankind intersects.

The presence of noise pollution can be highly correlated to the presence of environmental pollution as well.
For example, audio recordings can be used to determine the current volume of traffic in cities \citep{traffic_pollution_2015}. 
This information can be used to determine current air pollution levels, so that countermeasures can be taken accordingly (\eg directing traffic through alternative routes or issuing localised pollution warnings).
Aside from such short-term measures, acoustic monitoring can be used for long-term mitigation as well.
For example, monitoring of daily activity in modern megacities, in which significant transport results in a high degree of air pollution, can be used to inform public infrastructure projects. 
In addition, the emotional connotation of noise can be assessed automatically, providing the ability to monitor noise also in a qualitative rather than a sheer quantitative manner \citep{Schuller12-ARO}.

\subsection{Machines}
When it comes to machine monitoring in a globalised and modern world in which machines play a very important role, computer acoustics can help in machine management, making sure that machines operate at optimal power consumption levels, or identifying ruptures.
CA can assist in the timely identification of hazardous events and anomalous situations~\citep{ntalampiras10_interspeech, adaptiveframe, 5723010}.
Recent advances on using microphones and audio recordings to avoid adverse conditions for tools and machinery have shown promise~\citep{machinery_2020}. 
Problems can be solved by the utilisation of microphones in condition monitoring systems, which can be more beneficial than corrective maintenance since they allow early warnings of mechanical and electrical defects to prevent major component failures.
For example, acoustic monitoring has been successfully applied for monitoring bearing and gearboxes by using acoustic sensors~\citep{bearings_2019} and audio data collected by a cheap microphone has been utilised to monitor the condition of railway point machines~\citep{s16040549}.

Finally, there are several drawbacks to purely visual-based monitoring systems for industrial applications, such as occlusions by objects/smoke and illumination changes, to which audio-based detection systems would be more robust. 
In the study of \citet{industrial}, the proposed real-time acoustic anomaly detection system can be applied to recognise anomalous sound events such as fire, explosion, and glass breaking.
This can be substantially improved by the use of CA to expand the sensor coverage area, improve sensor robustness (\eg to low visibility conditions), and widen the scope of events that can be detected.

\section{Aether}
\emph{Aether} is the name given to the fifth element introduced by Plato and Aristotle, what came to be considered the quintessence of things.
It was an element that was considered discrete from the physical world, made of the air breathed by the Olympic gods.
It thus properly fits as a stand-in for AI, whose nature is primarily ethereal, in the sense that it consists of an agglomeration of algorithms and ideas implemented as software.

Coming up with a proper definition for AI is a difficult task, especially as the field itself is rapidly and constantly changing.
Perhaps the loosest of definitions is that it is a (computer) system which perceives and/or interacts with its environment in a way humans perceive as intelligent.
In a more practical sense, it is a system with a mechanism for incorporating, and later utilising, knowledge.
In recent years, knowledge acquisition has come to be dominated by the statistical learning paradigm~\citep{vapnik1999nature}, whose latest mainstay is deep learning~\citep{goodfellow2016deep}.

CA then is the sub-field of AI which concerns itself with auditory information.
This includes all aspects of intelligence which utilise sound as a communication medium.
The notion of communication is an important part of the definition, as communication theory is underpinned by the dual notions of sender and receiver.
A CA system can function as both: it may perceive sounds already present in its environment or transmit its own sounds into it. 

We note that `sound' here is not defined purely by anthropomorphic criteria.
While humans have come to define the gold standard as to what exactly constitutes intelligence, in the case of CA they are not the only, and perhaps not even the best, standard to measure up against.
With our limited hearing, human auditory abilities are dwarfed by those of several animals~\citep{wartzok1999marine}.
Accordingly, modern sound perception systems have already surpassed human capabilities, as the cases of hydrophones, ultrasound, and piezoelectric microphones demonstrate.
Likewise, sound production systems have now progressed to a state far superior than the human speech production system, and are able to both generate a much wider range of sounds and broadcast them further away than any human could.
Thus, we concern ourselves with sound as an umbrella term encompassing the propagation of vibrations over a diverse set of mediums.

However, capturing and transmitting sound is only the first step towards a cognitive CA agent.
Understanding what is being heard, deciding what to transmit in response, and synthesising it are much harder problems.
Sound generation, in particular for speech and music, has been revolutionised by WaveNet-derived architectures~\citep{oord2016wavenet, engel2017neural}.
Intelligent audio analysis has been accordingly shaped by the advent of deep learning~\citep{purwins2019deep}.
The bridge between sound generation and sound analysis has remained under-researched, at least as far as machine-to-animal communication is concerned, but can draw advances from recent developments in dialogue systems~\citep{chen2017survey} and reinforcement learning~\citep{sutton2018reinforcement}, both of which can serve as building blocks for designing an interactive agent.


\section{Audio Generation}
So far we have only dealt with audio in context of sensing for a (distributed) intelligence system whose solitary purpose is to detect events of interest in the surrounding environment.
However, an auditory communication channel can also be used for bidirectional communication between all entities connected to it.
This enables us to equip our intelligent system with a transmission module as well, thus converting it from a passive sensing module to an agent capable of interacting with its environment. This section can be regarded as an excursus to examples where an intelligent generation of audio might be beneficial for nature.

\subsection{Alarm Systems for Animals}
\label{sec:alarm}
Traditionally, the targeted receivers of machine communication have been human beings, as exemplified by the burgeoning field of human-computer interaction.
This is also true in the context of emergency systems; fire alarms, for instance, are used to notify humans to evacuate an area under threat. However, much of the negative effects of climate change are experienced by animals, not humans.
For example, the 2019 wildfires in southeast Australia have endangered several species by killing millions of individual animals~\citep{legge2020rapid}. 
Yet, thankfully, those fires left behind far less human casualties (34) as residents were evacuated from surrounding areas in timely fashion.
This indicates that part of those animal deaths could have been prevented if those animals had been appropriately evacuated as well.
Unfortunately, tracking, let alone evacuating, wild animals on such vast expanses of nature is currently at the limit of our technical capabilities.
Tracking has already been discussed in the previous sections. 
Now we will instead sketch a potential path towards a methodology suitable for animal evacuation.

The fundamental building block of such a system is machine-to-animal communication. 
Previous studies have established that certain animals use vocalisations to communicate with other member of their species, and in particular to alert them to the presence of threatening, or otherwise important, stimuli in their surrounding environment~\citep{bradbury1998principles}.
This existing mechanism could be appropriated by a distributed auditory intelligence to alert those animals, e.\,g., to the presence of a wildfire in their area.
From a hardware consideration, this would only entail the introduction of speakers alongside microphones; a minor addition in terms of costs.
From an AI perspective, this would require the transmission of appropriate vocalisations to be interpreted by the targeted animals as sounds of alarm, and correspondingly help them navigate the chaotic environment of a wildfire.

Generating those vocalisations is a challenging topic.
The vocabulary of animal communication has largely not been decoded, yet; thus, to the best of our knowledge, there is no available solution to this problem.
This is where the power of contemporary AI algorithms could come into play.
Reinforcement learning has long been established as a learning paradigm for agents that actively interact with their environment.
In this case, the generation of appropriate warning vocalisations could be recast as a reinforcement learning problem.
The system generates sounds and monitors the reaction of the animals in its surroundings, which in turn acts as a reward until the target reaction is achieved.

At this point, it is worth mentioning that such a system would not only be limited to the wildfire use case.
It can also be utilised for the protection of animals from other dangers, such as invasive hunter species or human poachers.
After detecting potential dangers, appropriate vocalisations would be produced to warn the animals in threat and help them evacuate the area.

\subsection{Open Space Active Noise Cancellation}
Noise pollution is a major environmental hazard which can harm humans and animals alike (\cf \cref{subsec:noise}).
The field which concerns itself with mitigating noise is noise cancellation, whereby measures are put in place to reduce, or even counteract, unwanted noise sources.
Traditionally, this is handled by passive measures, such as by introducing noise barriers in motorways, which requires a rough estimation of the properties of the noise expected at a given area.
In contrast, active noise cancellation (ANC) is based on an intentional transmission of a generated signal to destructively interfere with the noise that shall be cancelled.
With the advent of more intelligent audio analysis modules, open space ANC could be largely improved in the coming years.
In an ANC scenario, sensors and accompanying controllers are placed in several locations with the goal of identifying unwanted noise sources.
In turn, they inform an actuator which produces an `anti-noise' sound wave; essentially the same wave as the noise source but with an inverted phase~\citep{lam2021ten}.
Emphasis is therefore placed on the controller, which needs to decompose the input signal into its constituents so that the actuator can attempt to remove them, or at least reduce their unwanted side-effects.
Although this application has a large upside, its applicability remains limited (so far), as synthesising a proper sound wave to counteract specific sources in open spaces is a very challenging problem~\citep{srijomkwan2019cancellation}.
Should a complete cancellation appear too challenging, there also exist approaches to enhance the present audio composition and improve certain characteristics such as tonality, harmonicity, or tempo to increase human \citep{Baird20-IWT} (and potentially animal or plant) wellbeing.

\subsection{Sonification}
Different kind of data, thus also non-audio data, can be converted into an audio representation and radiated in form of audible sound to convey information. For example, the sonification of human movement pattern is a common practice in contemporary performing arts~\citep{landry2020interactive}, but can be also utilised as a feedback technique in physiotherapeutic treatment~\citep{guerra2020use}. Besides, data sonification can be a useful technique for applications in favour of nature and environment as well. In an exhibit at the National Center for Atmospheric Research, Boulder, Colorado, USA, climate data are presented as sounds in order to sensibilise visitors for global climate processes and to make them better understand the ongoing climate change. This innovative installation called ``Sounding Climate'' is based on data from the Community Earth System Model Large Ensemble Project and allows people to auditorily explore changes in temperature, precipitation, or sea ice over time since 1920~\citep{gardiner2018sounding}. A better understanding of ongoing climate processes might have a positive influence on people's attitude on our nature and their future interaction with our planet. Another project highlighting the potential of data sonification to raise people's awareness about unsustainable use of valuable natural resources, such as short water supplies, is the composition of ``The Lament of Las Tablas de Daimiel'' \citep{angeler2018sonifying}. This song builds upon a 71-year time series of rainfall and inundation area data and expresses the disruption of wetland in Spain due to agricultural transformation by means of a soprano and a bass voice. Intelligent sonification modules, \eg based on reinforcement learning techniques, could automatically identify, where and when people should get exposed to sounds generated from originally non-audible nature-related processes and which sounds/instruments should be used for specific settings and individuals to gain the maximum effect for the benefit of our planet.

\section{Discussion}
\label{Discussion}
The plenty of identified, already existing use cases of CA and audio generation in the context of environmental and climate-related questions, demonstrates the potential of this so far rather disregarded methodology to help saving our planet. Admittedly, most presented approaches do no directly contribute to `save the planet', but `just' to monitor nature or nature-related processes in the first place. However, monitoring of natural phenomena and early detection of changes or suboptimal environmental conditions represents an important requirement for any, audio- and non-audio-related, active technology-based intervention or initiation of human action. The presented overview shows that we are steering the right course by being open-minded for novel approaches at a time where our planet needs urgent help. However, what are the advantages of CA over alternative approaches? What are its limitations? Are there considerations from an ethics perspective? Is CA already set to reasonably contribute? In the remainder of this work, we aim to give answers to these questions. 

\subsection{Computer Audition vs Alternatives}
There are several ways to capture environmental and climate-related processes on our planet with audio recording being just one example suited for intelligent/machine learning-based analysis. Visual sensing is another example of established data collection for computer-based analysis, \ie computer vision. In addition, various other sensing alternatives, \eg other physical or chemical sensors, exist that are and might be used to gather input data for intelligent environment and climate monitoring and harm detection systems. Further, these sensors can be compared with the non-technology-based possibilities of the human body. 


\Cref{tab:comparison} compares these different modalities with audio in terms of data throughput, covered area, privacy, information richness, availability of ML models as well as costs and the degree to which they pollute the environment.

\begin{table}[t!]
  \caption{Comparison of acoustic sensing for environment and climate change monitoring vs other sensing modalities with regard to seven key criteria. Overall grading: +/green shading = advantageous, +-/orange shading = neutral, -/red shading = disadvantageous; m = meter(s); ML = machine learning} 
  \vspace{0.5cm}
  \label{tab:comparison}
  \centering
  \resizebox{\textwidth}{!}{%
  \begin{tabular}{>{\raggedright}p{0.16\textwidth} | >{\raggedright}p{0.16\textwidth} | >{\raggedright}p{0.16\textwidth} | >{\raggedright}p{0.16\textwidth} | >{\raggedright}p{0.16\textwidth} | >{\raggedright}p{0.16\textwidth}}
  \multirow{2}{*}{\diagbox[innerwidth=0.16\textwidth,height=7.1\line]{Criteria}{Modalities}} & Acoustic & Visual & Other physical & Chemical & Human \tabularnewline
   \multirow{2}{*}{} & \scriptsize Sensing of airborne sound including infrasound, sound in human audible frequency range, and ultrasound using microphones& \scriptsize Optical sensing by means of 2D, 3D, high-speed, thermal, aerial, and microscope cameras & \scriptsize Sensing of physical parameters other than airborne sound and visual information, such as temperature, current, humidity, fluid level, acceleration, pressure, and solid-borne sound & \scriptsize Sensing of chemical information, such as analyt composition, presence of specific elements, element concentration, and chemical activity & \scriptsize Non-technology-based sensing by means of human sight, hearing, smell, taste, and touch \tabularnewline
  \hline
  \hline
  Data throughput & \cellcolor{greenplus} + & \cellcolor{orangeplusminus} +- & \cellcolor{greenplus} + & \cellcolor{greenplus} + & \cellcolor{orangeplusminus} +-\tabularnewline
  \scriptsize Maximum amount of data that can be transmitted/processed per time & \cellcolor{greenplus} \footnotesize Extent of audio data is modest most of the time & \cellcolor{orangeplusminus} \footnotesize 2D images tend to be petite, while 3D images or videos tend to be comparatively large & \cellcolor{greenplus} \footnotesize Data stream is manageable in size & \cellcolor{greenplus} \footnotesize Data stream is manageable in size & \cellcolor{orangeplusminus} \footnotesize Sensing continuously; however, cannot really be transmitted from one human to another or to a computer \tabularnewline
  \hline
  Covered area & \cellcolor{greenplus} + & \cellcolor{greenplus} + & \cellcolor{orangeplusminus} +- & \cellcolor{orangeplusminus} +- & \cellcolor{orangeplusminus} +-\tabularnewline
  \scriptsize Sensor range regarding spatial coverage & \cellcolor{greenplus} \footnotesize Several hundred to several thousand m$^2$ (e.\,g., thunder); 360\textdegree\,recording angle possible (omnidirectional microphone) & \cellcolor{greenplus} \footnotesize Variable dependent on specific camera type; especially high range for aerial/satellite cameras; limited angle of view & \cellcolor{orangeplusminus} \footnotesize Very focused on one location; sometimes representative for a bigger area, \eg temperature & \cellcolor{orangeplusminus} \footnotesize Very focused on one location; sometimes representative for a bigger area, \eg soil composition & \cellcolor{orangeplusminus} \footnotesize Depends on the modality (vision, touch, etc.); can be large but also quite limited \tabularnewline
  \hline
  Privacy & \cellcolor{orangeplusminus} +- & \cellcolor{orangeplusminus} +- & \cellcolor{greenplus} + & \cellcolor{greenplus} + & \cellcolor{greenplus} + \tabularnewline
  \scriptsize Extent of personal data collected by the sensor; importance/possibility of data anonymisation \citep{mireshghallah2020privacy} & \cellcolor{orangeplusminus} \footnotesize Critical in case human voice is recorded \citep{kroger2019privacy}; not applicable for most scenarios in context of climate change  & \cellcolor{orangeplusminus} \footnotesize Critical in case human faces are recorded \citep{jose2019face}; not applicable for most scenarios in context of climate change & \cellcolor{greenplus} \footnotesize Not critical  & \cellcolor{greenplus} \footnotesize Critical blood or saliva analysis, or genetic sequencing not applicable in context of nature & \cellcolor{greenplus} \footnotesize Unproblematic as no human data are recorded at all \tabularnewline
  \hline
  Info richness & \cellcolor{greenplus} + & \cellcolor{greenplus} + & \cellcolor{orangeplusminus} +- & \cellcolor{orangeplusminus} +- & \cellcolor{greenplus} +\tabularnewline
  \scriptsize Amount of information extractable from recorded data & \cellcolor{greenplus} \footnotesize Many sound sources possible within one audio recording (e.\,g., animal sounds, rain, cars) & \cellcolor{greenplus} \footnotesize Many objects, classes, locations can be captured in one photo or video & \cellcolor{orangeplusminus} \footnotesize Mostly built to sense specific information, \ie only the desired information is recorded & \cellcolor{orangeplusminus} \footnotesize Mostly built to sense specific information, \ie only the desired information is recorded & \cellcolor{greenplus} \footnotesize Due to synchroneous multimodal sensing of the human body, lots of information is captured and processed in the brain \tabularnewline
  \hline
  ML models & \cellcolor{greenplus} + & \cellcolor{greenplus} + & \cellcolor{orangeplusminus} +- & \cellcolor{redminus} - & \cellcolor{redminus} -\tabularnewline
  \scriptsize Availability of (pre-trained) Machine Learning models & \cellcolor{greenplus} \footnotesize Many pretrained models available, which use raw audio or acoustic features as input & \cellcolor{greenplus} \footnotesize Many pretrained models available, which use raw video or visual features as input & \cellcolor{orangeplusminus} \footnotesize A few available models \citep{hanoon2021developing,wang2020towards,vamseekrishna2021prediction} & \cellcolor{redminus} \footnotesize Very few to none existent models with respect to climate & \cellcolor{redminus} \footnotesize No models available \tabularnewline
  \hline
  Costs & \cellcolor{greenplus} + & \cellcolor{orangeplusminus} +- & \cellcolor{greenplus} + & \cellcolor{orangeplusminus} + & \cellcolor{greenplus} +\tabularnewline
  \scriptsize Costs incurred by sensor & \cellcolor{greenplus} \footnotesize Microphones are very cheap; even specialised microphones are not too expensive & \cellcolor{orangeplusminus} \footnotesize Some cameras can be relatively cheap, others are very expensive; satellites or microcameras are extremely expensive & \cellcolor{greenplus} \footnotesize Mostly relatively cheap & \cellcolor{orangeplusminus} \footnotesize Sensor materials/resources can be expensive, \ie sensor costs vary a lot & \cellcolor{greenplus} \footnotesize The human body needs no further sensors; auxiliary means such as glasses are not too expensive \tabularnewline
  \hline
  Pollution & \cellcolor{greenplus} + & \cellcolor{orangeplusminus} +- & \cellcolor{greenplus} + & \cellcolor{orangeplusminus} +- & \cellcolor{greenplus} +\tabularnewline
  \scriptsize Pollution caused by sensor & \cellcolor{greenplus} \footnotesize Each sensor pollutes the environment to some degree; can be reused very often & \cellcolor{orangeplusminus} \footnotesize Shooting satellites into orbit emits lots of burnt gas and precipitates debris \citep{ross2018policy,adilov2015economic}
  & \cellcolor{greenplus} \footnotesize Each sensor pollutes the environment to some degree; can be reused quite often & \cellcolor{orangeplusminus} \footnotesize Each sensor pollutes the environment to some degree; can be reused only sometimes & \cellcolor{greenplus} \footnotesize Presumably the most environment-friendly option \tabularnewline
  \end{tabular}}
\end{table}

\subsection{Generalisation}
One of the major critiques of machine learning systems is their (lack of) ability to generalise.
This is most poignantly exemplified by their susceptibility to adversarial attacks~\citep{szegedy2014intriguing} -- minor perturbations to their inputs that lead to vastly different predictions.
While substantial research has been devoted to understand and mitigate this phenomenon, it is far from a solved problem, thus limiting the applicability of models in real world settings.
Moreover, according to the standard machine learning paradigm, the deployment environment must be identical to the training one; a constraint that is currently impossible to satisfy without vastly increasing the amount of data that is available for training.
This need is overcome via domain adaptation algorithms~\citep{ben2007analysis}, which explicitly minimise the discrepancy between source and target domains.

Moreover, the issue of generalisation is further complicated by the recently introduced notion of underspecification of machine learning architectures~\citep{d2020underspecification}, which also impacts auditory models~\citep{triantafyllopoulos2021fairness}.
Underspecification corresponds to the inability to predict the behaviour of a machine learning model in its deployment environment based on its behaviour in its training environment.
It constitutes a major challenge for real world applications as it makes model selection and, more crucially, model testing harder, leading to catastrophic failures during deployment.
This is of particular importance for a large-scale deployment required for environmental monitoring, where a model will have to perform equally well across several different locations, and, correspondingly, needs to be addressed via further research.

For audio analysis in particular, the notion of generalisation is closely linked to that of robustness to different perturbations, a topic that has received considerable attention over the years.
Traditionally, robustness has been studied under the auspices of speech enhancement~\citep{wang2018supervised, rethage2018wavenet, Liu19-NIT, triantafyllopoulos2019towards}, where (human) speech constitutes the signal of interest and (environmental) noise the unwanted interference that needs to be removed.
However, in our case the opposite would be true -- human voices would need to be removed in order to get more robust measurements of the environmental conditions~\citep{LiuTriantafyllopoulosRenetal}.

Overall, generalisation is still the primary concern, especially when the information procured by the deployed models will lead to decisions with important ramifications for the environment and our future.

\subsection{Efficiency}
Data is the fuel that drives contemporary AI applications.
Bigger and better data usually leads to bigger and better models, which require (vastly) more computational power to be trained~\citep{schwartz2020green}.
This trend places a significant, and ever increasing, strain on the world's resources~\citep{strubell2019energy}; a potentially flagrant oversight when creating models that attempt to reduce that strain.
Moreover, inconsiderate pursuit for more (sources of) data can potentially compromise privacy and democracy~\citep{yu2016big, helbing2019will}, both pillars of the social cohesion necessary to tackle climate change.

It is thus imperative to seek more resource efficient approaches for AI in general and CA in particular.
This includes the ability to learn from less data.
Two prominent learning paradigms that may be of use here are those of transfer learning~\citep{pan2009survey} and zero-shot learning~\citep{wang2019survey, xie2021zero}: the former corresponds to transferring knowledge from other tasks, usually ones for which data is abundant; the latter to generalising to classes unseen during training.
However, while both these paradigms have been successfully utilised in several applications, further work is needed to understand in more detail how they work and streamline their application~\citep{neyshabur2020being, triantafyllopoulos2021transfer}.

A key obstacle to the incorporation of more complex machine intelligence technology, such as deep learning-based solutions, into production-ready CA applications, is the significant amount of resources (\eg computational power) required for data collection, model training, and inference \citep{carbon_footprint_dl_2019}. 
As a result, a movement on \emph{sustainable AI} is being promoted among the research community. Sustainable AI can be understood as having two branches: \emph{AI for sustainability} and \emph{sustainability of AI}. The former one focuses on using AI towards the sustainable development goals\footnote{https://sdgs.un.org/es/goals}, and the latter aims at the reduction of carbon emissions and computing power used in AI methodology itself \citep{sustainable_ai_2021}.
The whole research community is slowly but steadily detouring towards changing the entire lifecycle of AI products (\ie idea generation, training, and implementation) to achieve greater ecological integrity. It is focusing its efforts on different alternatives, such as the use of flexible and lightweight data augmentation, which can be incorporated into training processes to boost performance and robustness, saving training resources and improving generalisation of models. Besides cross-corpus learning, cross-modal transfer learning techniques can be used to leverage knowledge from pre-trained neural networks based on state-of-the-art architectures across different domains. Furthermore, novel technologies for resource optimisation, such as low-parameter architectures~\citep{amiriparian2021deepspectrumlite}, leveraging compression techniques~\citep{cheng2018model}, pruning, or teacher student networks,
are being further developed to reduce the computational footprint of AI.
Sustainable AI is about developing AI that is compatible with sustaining environmental resources for current and future generations, while keeping in mind that there are environmental costs to AI itself. 

\subsection{Trustability}
While AI is often touted as a technology capable of operating completely autonomously, the truth is that most AI applications are embedded in an ecosystem involving other entities, primarily humans.
This introduces the additional requirement of trustability: The entities that interact with an AI program must trust the program under consideration of known limitations, in order to take its outputs and decisions properly into account.
For humans, trustability is largely related to explainability~\citep{ignatiev2020towards}, a concept that remains elusive~\citep{lipton2018mythos} but on which significant progress has nevertheless been made~\citep{adadi2018peeking}.
Explainability corresponds to the ability of an AI agent to \emph{explain} its decisions, that is, break its causes down to easily digestible attributions.
Interestingly, in our application scenario, it is possible to imagine other entities coming in contact with AI models; in particular, animals (\cf \cref{sec:alarm}).
Consequently, these entities should also be taken into account when attempting to endow any CA systems with notions of explainability and trustability.
While this may seem as an exotic field of study (that to our knowledge remains unresearched), it may prove crucial for saving the world's biodiversity.

\subsection{Bias}
\label{sec:bias}
All real-world AI technologies are plagued by bias. Audition in particular is affected, as sources of signal are difficult to determine and can come from a wide range of generative factors. The typical AI setup involves sampling from the real world, creating a dataset, and training a model to perform a task in this sampled space. Through identifying spurious correlations and bias from study design sampling methods, AI models are often able to perform well in these sampled datasets. This wrongly suggests a true underlying signal and so opportunity for AI to help in a field. This has been demonstrated in the recent COVID-19 pandemic, where early research suggested COVID-19 was uniquely identifiable from infected individuals' respiratory sounds \citep{coppock2021SevenGrainsofSalt}. Moving forward, CA for the environment should be directed down biologically plausible avenues and significant work should be done on study design and bias mitigation.

\subsection{Ethical Considerations}
\label{Ethics}
Over the years, several different guidelines have been proposed, each encompassing a different set of principles for ethical AI~\citep{jobin2019global, hagendorff2020ethics, Quadrianto21-EEM}, 
including such tailored for specific audio \citep{Batliner20-EAG}, 
all of which are relevant for CA systems put in use against climate change.
We highlight two of the most pertinent ones below, namely \emph{fairness} and \emph{privacy}, and refer the interested reader to recent overviews of ethical AI for a more general discussion

Fairness is one important consideration for the ethical application of AI.
It is necessary to provide adequate performance guarantees for all parts of our planet's ecosystem, irrespective of where in the world those may be~\citep{triantafyllopoulos2021fairness}.
This becomes particularly pressing as the areas of the world that are most in danger tend to be more strained for resources.
It is thus easy to imagine a scenario where the richest countries collect most of the data within their borders, leading to an underrepresentation of the world's more vulnerable countries in the training, and, consequently, to the underperformance of the algorithms when deployed on their premises.

Privacy is another big risk in the era of big data~\citep{yu2016big}.
This regards the possibility of mass surveillance, especially in applications such as environmental monitoring, where the large-scale deployment of several sensors is mandated.
Such a widespread deployment of sensors, in our case microphones, puts in danger citizens around the world whose data might be stored, shared, and analysed without their knowledge or permission.
Protecting that privacy is possible under several different paradigms.
Perhaps the most straightforward one is to remove all speech information immediately after capture and before any subsequent processing is done~\citep{LiuTriantafyllopoulosRenetal}; this strategy, however, might lead to suboptimal results as any source separation algorithm invariably introduces unwanted artefacts and information loss.
Alternatively, the federated learning paradigm could be followed, where the data is processed only at edge nodes and that information is then streamed in order to train a global model while preserving the privacy of those nodes~\citep{9084352}. Federated learning has been successfully applied in many areas, such as healthcare~\citep{xu2021federated}, sensing in smart cities~\citep{s20216230}, and financial services~\citep{9346395, Long2020}. However, there are core challenges of federated learning~\citep{9084352}. For instance, this paradigm is not completely privacy-proof~\citep{bonawitz2017practical} and may also lead to worse performance~\citep{bonawitz2017practical, mcmahan2018learning}.
Altogether, privacy remains a largely unsolved problem for AI systems operating on data that may include personal information.
Nevertheless, for many of the applications outlined here, this is not a primary concern, as they focus on wildlife monitoring.

Finally, we want to point out that the typical formulation of ethical dilemmas in AI is highly anthropocentric, thus, concentrating primarily on the effect of AI on human affairs, and, even in the cases where it touches upon its ecological aspects, it does so from the human perspective as well.
This anthropocentric view has been challenged by several scholars arguing for the development of environmental or ecological ethics~\citep{brennan2002environmental, curry2011ecological}.

At this point, we have to generally recap that the ultimate aim of this work was to give a current overview of applications where CA (and intelligent audio generation) technology might contribute to `save our planet'. Thus, we need to shed light on the essential question of what does saving our planet actually mean. Strictly speaking, saving our planet implies that the Earth is preserved as long as possible, with or without a human population. Nevertheless, some applications outlined in this work seem to be beneficial for human individuals at first glance rather than for the planet itself, such as the early detection of a tornado. However, in this work, we regard the basic need of mankind to further populate the Earth as its habitat as given and, under these circumstances, the prevention of human beings from physical harm or material damage might prevent costly and pollutive medical interventions or the resources-consuming replacement of goods. Thereby, we followed the dominant paradigm, putting humanity and its interests in the centre of discussion, but we acknowledge the need for a holistic consideration of our natural ecosystem.

\subsection{Outlook}
\label{Outlook}
CA is already used in a variety of applications in the context of environmental and climate change-related issues. Compared to other alternatives, CA has several advantages, such as providing cost-effective coverage of large areas and enabling early detection of changes even before they become apparent in other modalities. Therefore, CA seems to be on the right track and definitely has the potential to impact future applications. However, more work needs to be done to exploit the full potential of audio intelligence in this area. In particular, it is important to overcome some methodological limitations, while always respecting the ethical framework by, e.\,g., ensuring privacy and fairness of the models. In order to further increase the power of audio intelligence applications, a combination of CA with other techniques would be worth striving for in the future, leading to intelligent multimodal sensing applications, e.\,g., in the form of computer audio-vision based systems. 

With this work we would like to convince researchers of the potential of audio intelligence in combating climate change and hope that many will follow our call-to-action to continue what has already been initiated and explore new applications of CA and intelligent audio generation in the various fields related to environmental and climate issues in order to jointly contribute to the preservation of our habitat and to saving our planet.

Our planet is crying for help. Let us use Computer Audition to hear those cries and translate them to action!

\section*{Acknowledgements}

This work was funded by the German Research Foundation (DFG; Reinhart Koselleck project, No.~442218748: AUDI0NOMOUS). The authors want to thank everyone who takes responsibility for our beautiful planet and makes an effort to prevent nature from harm.  
Zero GPU hours were consumed to provide the contents discussed herein. 

\section*{Conflict of Interest Statement}
The author Bj\"orn W.\ Schuller was employed by the company audEERING GmbH. The remaining authors declare that the research was conducted in the absence of any commercial or financial relationships that could be construed as a potential conflict of interest.

\bibliographystyle{plainnat}
\bibliography{mybib}
\end{document}